\documentclass[useAMS]{mn2e}

\newcommand{\be} {\begin{equation}}

\newcommand{\ee} {\end{equation}}

\newcommand{\CXO}{{\em Chandra}}
\newcommand{\bc}{\begin{center}}
\newcommand{\ec}{\end{center}}

\def\ltsima{$\; \buildrel < \over \sim \;$}
\def\lsim{\lower.5ex\hbox{\ltsima}}
\def\loe{\lower.5ex\hbox{\ltsima}}
\def\gtsima{$\; \buildrel > \over \sim \;$}
\def\gsim{\lower.5ex\hbox{\gtsima}}
\def\goe{\lower.5ex\hbox{\gtsima}}

\def \cm2{cm$^{2}$\,}

\def\ergs {erg\,s$^{-1}$}
\def\ergscm2 {erg\,s$^{-1}$cm$^{-2}$}

\def\rra {RRAT\,J1819--1458\,}
\def\rrb{RRAT\,J1913+1333\,}
\def\rrc {RRAT\,J1317--5759\,}

\begin{document}

\title[ESO-VLT observations of \rra\, and \rrc]{Near-infrared observations of Rotating Radio Transients}

\author[Rea et al.]{N. Rea$^{1,2}$\thanks{Ramon y Cajal Research Fellow; rea@ieec.uab.es.}, G. Lo Curto$^{3}$, V. Testa$^{4}$, G. L. Israel$^{4}$, A. Possenti$^{5}$, M. McLaughlin$^{6,7}$, \newauthor F. Camilo$^{8}$, B.M. Gaensler$^{9}$, M. Burgay$^{5}$ \\
$^{1}$ Institut de Ciencies de l'Espai (ICE-CSIC, IEEC), Campus UAB, Fac. de Ciencies, Torre C5-parell, 2a planta, 08193 Barcelona, Spain \\
$^{2}$ University of Amsterdam, Astronomical Institute ``Anton Pannekoek'', Postbus 94249, 1090 GE, Amsterdam, The Netherlands \\
$^{3}$ European Southern Observatory, Av. Alonso de Cordova 3107, Vitacura, Santiago, Chile \\
$^{4}$ INAF - Osservatorio Astronomico di Roma, Via Frascati 33, 00040 Monte Porzio Catone, Italy\\
$^{5}$ INAF - Osservatorio astronomico di Cagliari, Poggio dei Pini, Strada 54, 09012 Capoterra (CA), Italy \\
$^{6}$ Department of Physics, West Virginia University, Morgantown, WV 26506, USA \\
$^{7}$ National Radio Astronomy Observatory, Green Bank, WV 24944, USA \\
$^{8}$ Columbia Astrophysics Laboratory, Columbia University, New York, NY 10027, USA \\
$^{9}$ Sydney Institute for Astronomy, School of Physics, The University of Sydney, NSW 2006, Australia}
\input psfig.sty

\pagerange{\pageref{firstpage}--\pageref{lastpage}} \pubyear{2009}

\maketitle

\label{firstpage}

\begin{abstract}

  We report on the first near-infrared observations obtained for Rotating RAdio Transients (RRATs). Using adaptive optics devices
  mounted on the ESO Very Large Telescope (VLT), we observed two
  objects of this class: \rra\, and \rrc. These observations have been
  performed in 2006 and 2008, in the J, H and K$_{\rm s}$ bands. We
  found no candidate infrared counterpart to \rrc\, down to a limiting
  magnitude of K$_{\rm s} \sim 21$. On the other hand, we found a
  possible candidate counterpart for \rra\, having a magnitude of
  K$_{\rm s}=20.96\pm0.10$. In particular, this is the only source
  within a 1$\sigma$ error circle around the source's accurate X-ray
  position, although given the crowded field we cannot exclude that
  this is due to a chance coincidence. The infrared flux of the putative 
  counterpart to the highly magnetic \rra\, is higher than expected from a normal radio
  pulsar, but consistent with that seen from magnetars. We also
  searched for the near-infrared counterpart to the X-ray diffuse
  emission recently discovered around \rra, but we did not detect this
  component in the near-infrared band. We discuss the luminosity of the
  putative counterpart to \rra\, in comparison with the near-infrared
  emission of all isolated neutron stars detected to date in this band
  (5 pulsars and 7 magnetars). 
  
  \end{abstract}


\section{Introduction}

Rotating RAdio Transients (RRATs) are a recently discovered class of
neutron stars (McLaughlin et al.~2006, 2009; Keane et al. 2010)
characterized by dispersed radio bursts with flux densities (at a
wavelength of 20~cm) ranging from $\sim$100 mJy to 4 Jy, durations from 2
and 30 ms, and average intervals between repetition from 4 minutes to
3 hours. Periodicities have been inferred through the study of arrival
times of these bursts, ranging between 0.1--7\,s. The timing solutions
derived from the radio bursts, i.e. their periods and period
derivatives, indicate that they are
neutron stars  (McLaughlin et al.~2006, 2009).


\begin{table*}
\begin{minipage}{150mm}
\centering{
\begin{tabular}{cccccc}
\hline
\hline
\multicolumn{6}{c}{RRAT\,J1819-1458} \\
\hline
Date & Filter & FWHM(pix) & Exp.Time & Mag. Lim.(3$\sigma$) & zero point \tabularnewline
\hline
 2006 Jun. 26 & K$_{\rm s}$ & 2.9  & 2280 & 21.4  & 22.52$\pm$0.06 \tabularnewline
 2006 Jun. 26 & J & 5.0  & 1620 & 22.8  & 23.27$\pm$0.18 \tabularnewline
 2006 Jun. 28 & H & 3.2  & 1620 & 21.8 & 23.12$\pm$0.07 \tabularnewline
 2006 Jul. 21 & H & 3.1 & 1800 & 22.4  & 23.43$\pm$0.07 \tabularnewline
 2006 Jul. 31 & J & 10.0 & 1800 & 22.9  & 23.67$\pm$0.06 \tabularnewline
 2008 Jul. 18 & K$_{\rm s}$ & 4.0 & 1080 & 20.7 &  22.40$\pm$0.08 \tabularnewline
\hline
\hline
\multicolumn{6}{c}{RRAT\,J1317-5759} \\
\hline
Filter & Date & FWHM(pix) & Exp.Time & Mag. Lim.(3$\sigma$) & zero point \tabularnewline
\hline
2006 Apr. 02 & K$_{\rm s}$ & 3.2 & 2280 &  20.90 &  22.48$\pm$0.06 \tabularnewline
2006 Apr. 02 & J & 6.2 & 2520 &  22.70  & 23.14$\pm$0.07 \tabularnewline
2006 Apr. 27 & H & 3.9 & 1320 &  20.90  & 22.55$\pm$0.06 \tabularnewline
2006 Apr. 28 & H & 3.9 & 2280 &  22.80  & 23.06$\pm$0.06 \tabularnewline

\hline
\hline
\end{tabular}
\label{tabobs}
\caption{Summary of the ESO-VLT observations of \rra\, and \rrc .  The NACO pixel size corresponds to 0.027\arcsec .}}
\end{minipage}
\end{table*} 

The detection of such sources is rather difficult, mainly due to the
very tiny duty cycle of their radio bursts (0.1--1\,s of radio
emission per day). Detailed population studies show that there may be more RRATs than canonical rotational powered radio pulsars in our Galaxy (McLaughlin et al.~2006, 2009; Keane \& Kramer 2008).

Determining the nature of the emission from these objects and how many
RRATs are present in our Galaxy is of paramount importance for pulsar
emission theories, as well as for neutron star population studies. Up
to now many hypotheses on the nature of these objects have been put
forward, based on the comparison with the radio pulsar class:
e.g. that the RRATs may be neutron stars near the radio ``death line''
(Zhang et al. 2006), or that the sporadicity of the RRATs is due to
the presence of a circumstellar asteroid belt (Cordes \& Shannon~2008;
Li~2006) or a radiation belt such as seen in planetary magnetospheres
(Luo \& Melrose~2007). Note that, even though they are all
characterized by a similar radio bursting activity, RRATs might be an
inhomogeneous group of sources, comprising neutron stars belonging to
different classes. Interesting possibilities are the
connection with i) the transient magnetars, given the recent radio
detection during their outbursts (Camilo et al.~2006, 2007); ii) with
the X-ray Dim Isolated Neutron Stars (XDINSs; Haberl 2007; van Kerkwijk
\& Kaplan 2007; Popov 2006), given the thermal X-ray emission discovered for \rra,
being so far the most magnetic RRAT (Reynolds et al. 2006; McLaughlin
et al. 2007); and iii) with the radio pulsars showing thermal X-ray emission, such
as PSR~B0656+14, which has been argued to be a close RRAT (Weltevrede
et al. 2006).

Optical and infrared counterparts have been identified so far only for
ten rotational powered radio pulsars (V$\sim 26$; see Mignani et
al. 2009 for a recent review), four XDINSs (V$\sim 25$; van Kerkwijk \& Kaplan 2007), and seven magnetars (K$_{\rm
  s}\sim$20; Israel et al. 2004; Testa et al. 2008; Mereghetti
2008). Despite being all isolated neutron stars, their optical and
infrared emission has distinctive properties depending on the
source class. Very little is known about a possible optical or infrared component
in the emission of RRATs, beside the non-detection of optical bursting
emission (Dhillon et al. 2006). The detection and characterization of
the nature of the putative infrared emission of RRATs is a crucial
help to assess their relation with any of the aforementioned neutron
star classes.

We report in this paper on the first infrared observations of RRATs, in particular we study two sources of this class having a relatively good positional accuracy: \rra\, and
\rrc\footnote{Within our ESO--VLT observing campaign, we also took observations 
of \rrb. However during the submission of this paper, we became aware
  that the published position of this RRAT was off by $\sim$3\arcmin ,
  hence the correct position (McLaughlin et al. 2009) lies outside the VLT--NACO field of view
  of our observations.}. The paper is structured as follows: in
\S\,\ref{data} we report on the observations and data analysis, in
\S\,\ref{results} we show the results of our infrared study, and 
discussion follows (\S\,\ref{discussion}).


\begin{figure*}
\centering{
\hspace{0.5cm}
\psfig{figure=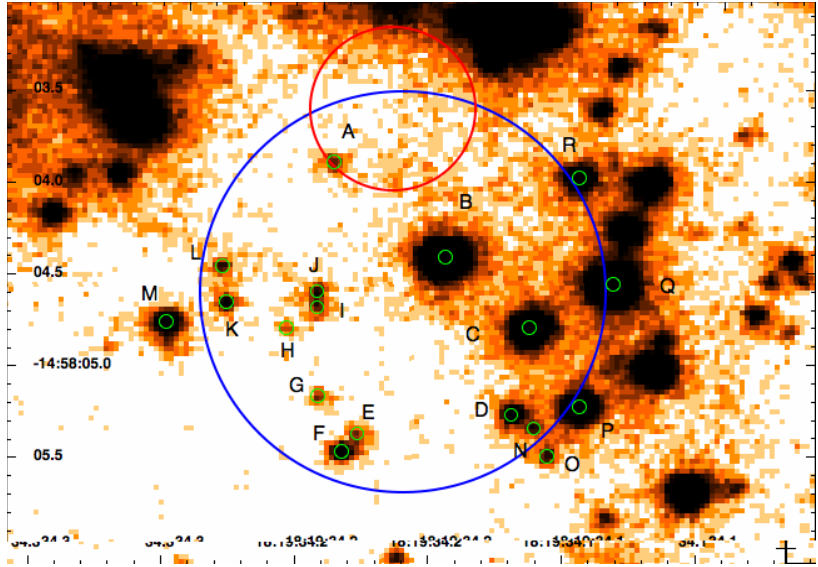,width=14cm,angle=270}}
\caption{\rra: ESO--VLT K$_{\rm s}$-band image of the June 2006 observation. The 1$\sigma$ error circle around the new X-ray position of the RRAT is marked (0\farcs45 ; Rea et al. 2009a) . Small circles (see also Tab.\,2) show the position of the stars close or near the RRAT 2$\sigma$ old X-ray position (1\arcsec ; Reynolds et al. 2006). All positional errors are weighted with the uncertainty in the NACO astrometry (see text for details), and the coordinate grid in RA and Dec (J2000) is overplotted. North is on top and East is left. } 
\label{map1819}
\end{figure*}

\begin{table*}
\begin{tabular}{lll|cc|cc|cc}
\hline
\hline
\multicolumn{3}{c}{Source} & \multicolumn{2}{c}{J band} & \multicolumn{2}{c}{H band} & \multicolumn{2}{c}{K$_{\rm s}$ band} \tabularnewline
\hline
ID & RA (h m s) & Dec (d m s) & 2006-06-26 & 2006-07-31 & 2006-06-28 & 2006-07-21 & 2006-06-26 & 2008-07-18   \tabularnewline
\hline
A  & 18 19 34.20 & -14 58 03.9 &     -	&	-	&	-	&	- & 20.96$\pm$0.10	      & -                      \tabularnewline
B & 18 19 34.15 &  -14 58 04.4 &  21.05$\pm$0.19 &	21.03$\pm$0.07 &	18.56$\pm$0.07	& 18.52$\pm$0.06 & 17.89$\pm$0.06	& 17.95$\pm$0.08       \tabularnewline
C & 18 19 34.12 & -14 58 04.8 &   22.52$\pm$0.21 & 22.89$\pm$0.12 &	19.43$\pm$0.09	& 19.49$\pm$0.06 & 18.49$\pm$0.07	& 18.47$\pm$0.09  \tabularnewline
D &  18 19 34.13	&  -14 58 05.3    &   22.69$\pm$0.21	& 22.40$\pm$0.10& 	20.50$\pm$0.10	& 20.51$\pm$0.07 & 19.99$\pm$0.08	& 20.10$\pm$0.10 \tabularnewline
E &  18 19 34.19 &	 -14 58 05.4  &    -	&	-	&	-	&	- & 20.94$\pm$0.10	& -                    \tabularnewline
F  & 18 19 34.19 &	 -14 58 05.5   &   -	& 	-		& 21.83$\pm$0.20	& 21.36$\pm$0.08 & 20.13$\pm$0.08	& 20.35$\pm$0.10 \tabularnewline
G & 18 19 34.20	&  -14 58 05.2   &  -	&	-	& 	-		& 22.49$\pm$0.11  & 20.87$\pm$0.10	& - \tabularnewline
H & 18 19 34.21 &	 -14 58 04.8    &   -	&	-	&	-	&	-  & 21.09$\pm$0.11	& - \tabularnewline
I  & 18 19 34.20	&  -14 58 04.7  &  -	&	-	&	-	&	- & 20.97$\pm$0.10 & 	- \tabularnewline
J & 18 19 34.20 &	 -14 58 04.6   &  -	&	-	&	-	&	21.86$\pm$0.09 &  20.86$\pm$0.10 &  	- \tabularnewline
K & 18 19 34.24 &	 -14 58 04.7   &  -	&	-	&	21.24$\pm$0.13 & 21.10$\pm$0.08 & 20.61$\pm$0.09	& 20.59$\pm$0.11 \tabularnewline
L & 18 19 34.24	&  -14 58 04.5   & -	&	-	&	-	& 	- & 20.64$\pm$0.09 &	-  \tabularnewline
M$^*$ & 18 19 34.26	& -14 58 04.8  &   23.10$\pm$0.22	& -	          &	20.19$\pm$0.10 &	20.20$\pm$0.07 & 19.47$\pm$0.07 & 19.45$\pm$0.09 \tabularnewline
N & 18 19 34.12	&  -14 58 05.3  & -	&	-& 		-	&	-  & 20.90$\pm$0.10&	- \tabularnewline
O & 18 19 34.12	&  -14 58 05.5    &   -	&	-& 		21.38$\pm$0.14 &	21.59$\pm$0.08 &  20.83$\pm$0.10 &	- \tabularnewline
P &  18 19 34.10 &	 -14 58 05.2 &  21.89$\pm$0.19 & 	21.82$\pm$0.08& 	19.37$\pm$0.07 &	19.36$\pm$0.06 & 18.89$\pm$0.07	& 18.87$\pm$0.09 \tabularnewline
Q & 18 19 34.09	&  -14 58 04.6   &  21.56$\pm$0.19& 	21.55$\pm$0.08& 	18.74$\pm$0.07 &	18.76$\pm$0.06 &  17.90$\pm$0.06 &	17.82$\pm$0.08 \tabularnewline
R & 18 19 34.10	&  -14 58 04.0   &   22.80$\pm$0.21& 	22.59$\pm$0.10& 	19.87$\pm$0.09 &	19.91$\pm$0.07 & 19.37$\pm$0.07 &	19.20$\pm$0.09 \tabularnewline

\hline
\hline
\end{tabular}
\label{rrattab}
\caption{Magnitudes of the sources close to \rra\, (see also Fig.\,1). RA and Dec refer to J2000. $^*$ Possible extended source. }
\end{table*} 

\section{Observations, Calibrations and Analysis}
\label{data}

\begin{figure}
\centerline{\psfig{figure=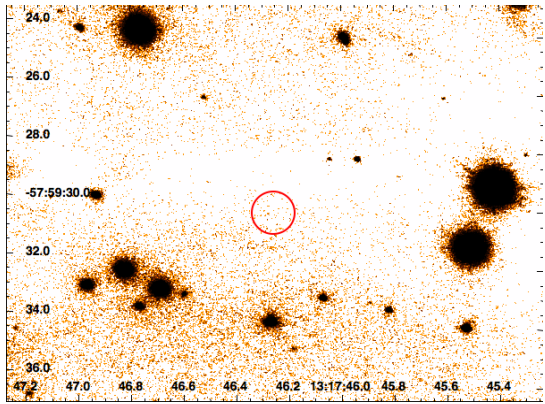,width=10cm,angle=270}}
\caption{ \rrc: ESO--VLT K$_{\rm s}$-band image with the 1$\sigma$ error circle around the refined
  position (see text for details).  The coordinate grid in RA and Dec (J2000) is overplotted. North is
  on top and East is left. }
\label{map1317}
\end{figure}

Observations have been obtained during years 2006 and 2008, with the
ESO VLT-UT4 Yepun, equipped with the adaptive optics near-infrared camera
NACO (NAOS/CONICA) in three filters: J ($\lambda = 12650 \AA$,
FWHM = 2500 $\AA$), H ($\lambda = 16600 \AA$, FWHM= 3300 $\AA$), and K$_{\rm s}$ ($\lambda = 21800 \AA$, FWHM =
3500 $\AA$). In some cases the objects have been observed in more than
one epoch, as summarized in the Tab.\,\ref{tabobs}. All the
observations have been performed with airmass between 1.0-1.3, a seeing $<$1\farcs5, and using natural guide stars for the adaptive optic system. \\
\indent
The data were reduced using the standard strategies for near-infrared data reduction. Flat fielding was done via internal lamp flats. Sky emission has been computed using IRAF\footnote{IRAF is
distributed by the National Optical Astronomy Observatories, which are
operated by the Association of Universities for Research in Astronomy Inc., under cooperative agreement with the National Science
Foundation.}  on the dithered images and then subtracted on each image before de-jittering. We used the \textit{eclipse}\footnote{www.eso.org/eclipse} package to co-add the images. Every pointing is the co-addition of three  images of 40s - 60s each, depending on the filter. This co-addition is performed by the computer on board the instrument, and we obtain only one image per dithering position. Fig.\,\ref{map1819} and Fig.\,\ref{map1317} show
finding charts in the K$_{\rm s}$ band for the two studied RRATs, with
the source positions highlighted.\\
\indent
We first reduced the images with the IRAF version of \textit{daophot}, using the PSF fitting
algorithm. Aperture photometry was also performed using an aperture
diameter twice the measured FWHM.  Aperture and PSF photometries have
been compared by applying aperture corrections using the IRAF
\textit{daogrow}  algorithm, which  calculates the  aperture correction
extrapolating to an aperture at infinity. \\
\indent
Furthermore, due to the crowding of the fields, we checked the photometry via a
deconvolution code: DECPHOT (Gillon et al. 2006, 2007; Magain et
al. 2007), based on the "MCS" algorithm (Magain, Courbin \& Sohy
1998).  The basic idea of the MCS code, is that of "partial"
deconvolution: the data are deconvolved to a higher resolution,
but at a level that the Nyquist criterium for a correct sampling was satisfied. This resulted in excellent photometric and astrometric
accuracy.  At variance with the standard MCS method, the code by
Gillon et al. (2006, 2007) is faster, determines the PSF "on the fly" and fits the sky
background.  We report hereafter the photometry delivered by this
code, which was compatible, for the most isolated objects, with the photometry
from the {\tt imexam} task of IRAF and with {\tt daophot}.\\
\indent
Calibrations have been performed using photometric standard stars observed within 4 hours from our target. The calibrated catalogs were then matched for each filter, and compared to verify the self-consistency of the
calibrations. Small differences in the calibrations have been found
between the J, H, and K$_{\rm s}$ images in different epochs, in this case
the best quality epoch has been chosen as reference and the magnitudes
of the other epoch reported to it, for each source respectively.\\
\indent
Further check with the 2MASS catalog has been performed but
unfortunately the 2MASS objects present on the images were saturated
or at the saturation limit and thus the zero point check with the
2MASS was not constructive. Furthermore, the 2MASS catalog was built
from images with a 2$\arcsec$ pixel size, and in many cases the
measured magnitudes correspond to blended objects which are, instead,
resolved in our higher resolution images. The calibrated catalogs for the single filters were then matched to produce a final master catalog. \\
\indent
In all cases, astrometric  calibration was performed using 2MASS stars
as  reference, yielding  an  average uncertainty  of $\sim$ 0\farcs33 (1$\sigma$ confidence level), after  accounting  for the  intrinsic  astrometric  accuracy of  2MASS and the  r.m.s. of the astrometric fits.  \\
\indent


\begin{figure*}
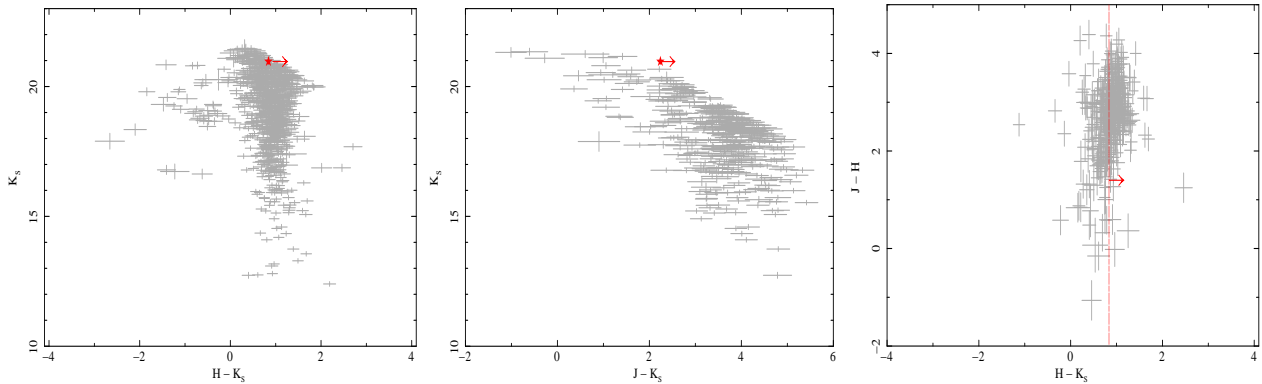

\centering{
\hbox{
\psfig{figure=1819_2006_HminusKsvsKs.ps,width=5.5cm,height=5cm,angle=270}
\psfig{figure=1819_2006_JminusKsvsKs.ps,width=5.5cm,height=5cm,angle=270}
\psfig{figure=1819_2006_HminusKsvsJminusH.ps,width=5.5cm,height=6.4cm,angle=270} }
\caption{From left to right: color-magnitude (H--K$_{\rm s}$,K$_{\rm
    s}$; J--K$_{\rm s}$, K$_{\rm s}$) and color-color diagrams
  (J-K$_{\rm s}$, H-K$_{\rm s}$,) of the VLT--NACO imaged area of
  \rra\, taken in year 2006 (see also Fig.\,1 and Tab.\,2). The star and
  vertical line are relative to the limits we have for our candidate
  counterpart, source A.}}
\label{cmd}
\end{figure*}


\section{Results}
\label{results}

\subsection{\rra}

This is the most prolific radio burster among the RRAT class, with a
magnetar-like dipolar magnetic field at surface of
$B\sim5\times10^{13}$\,Gauss, and a spin period of $\sim$4.26\,s
(McLaughlin et al. 2006; Lyne et al. 2009). It is so far the only
member of the class for which X-ray emission has been discovered
(Reynolds et al. 2006; McLaughlin et al. 2007), with a luminosity of
L$_X=4\times10^{33}$\ergs\, (assuming a distance of 3.6\,kpc). Diffuse X-ray emission has been recently reported for \rra\, with a
luminosity of $\sim10^{32}$\ergs\, and extending until $\sim$13\arcsec
from the source (Rea et al. 2009a). Furthermore, the RRAT error circle has
been recently refined performing a bore-site correction of a new
\CXO\, observation which were tied to the 2MASS field using a field
star present in both the X-ray and infrared images. This resulted in
an accurate position of RA 18:19:34.173, Dec -14:58:03.57 (J2000) with
a 1$\sigma$ error of $\sim$0\farcs3 (Rea et al.~2009a).

We took observations in the three standard near-infrared bands (J, H
and K$_{\rm s}$), and in two epochs per filter (see Tab.\,\ref{tabobs}). These
resulted in the identification of a possible candidate near-infrared
counterpart to \rra, the only source within the 1$\sigma$ X-ray positional error circle (our source A; see Fig.\,\ref{map1819} and Tab.\,2;  note that we took into account the astrometric error reported in \S\ref{data}, inferring a final 1$\sigma$ error circle of 0\farcs45). Although a few other sources lies close to \rra, in particular if we enlarge the X-ray
positional error circle to e.g. 90\%, they are all rather bright
objects, not compatible with the expected magnitudes from an isolated
neutron star of any kind. \\
\indent Within the limiting magnitude of our deepest exposure, K$_{\rm
  s}=21.4$ (see Tab.\,1), we significantly detected 380 sources within the calibrated
NACO field of 25\arcsec$\times$22\arcsec, which means an average of 1
source on an area of $1\farcs45^2$. Given such a crowded
region we cannot exclude that the identification of source A as a
possible counterpart through its proximity to \rra's X-ray position
might be due to a chance coincidence, the probability of which is $\sim$43\%. However, as a further selection criteria, assuming that an isolated neutron star at 3.6\,kpc is not expected to have a magnitude $<$19 (which is the brightest K magnitude observed in an isolated neutron star, namely for SGR\,0501+4561 and SGR\,1806--20; Levan et al. 2010 in prep.; Israel et al. 2005), we are then left with  115 sources in the field responding to such selection, hence the probability of having in coincidence with the X-ray position a source with a magnitude expected for a possible counterpart is $\sim$14\%.

The source is very faint and not visible in H and J bands within our
limiting magnitudes (see Tab.\,2). We detect the source only in our
deepest K$_{rm s}$ image on June 2006, however no variability can be claimed
since the following epoch was not deep enough to detect such a
faint source. We did not find any J, H and K$_{\rm s}$ variability for
the objects close to our target, both investigating different frames
of a single observation, and comparing the several epochs.  In Fig.\,3 we show the color-magnitude and color-color diagrams for all sources in
the field of \rra. We did not find any source with peculiar colors. We
over-plot the limits we have for our candidate counterpart (source A),
which unfortunately was detected only in the K$_{\rm s}$ band, hence we have only limits on its colors, namely J-K$_{\rm s}>$1.94 and H-K$_{\rm s}>$1.44 .  If source A is indeed the counterpart to \rra, then its K$_{\rm s}$ luminosity would be L$_{\rm NIR}=1.62\times10^{30}$\ergs\, (assuming a 3.6\,kpc distance; see also Tab.\,3), while if it is not, the 3$\sigma$ upper limit we derived corresponds to L$_{\rm NIR}<1.51\times10^{30}$\ergs .

Furthermore, we searched for the possible near-infrared counterpart to
the X-ray diffuse emission discovered around \rra, removing all the
sources in the field on the basis of their fitted PSF.  We did not find
any extended emission down to a limiting K$_{\rm s}$ surface brightness of $> 18.5$ mag/arcsec$^2$ (at a 3$\sigma$ confidence level).



\subsection{\rrc}

With a period of $\sim$2.64\,s and a magnetic field of
$\sim5\times10^{12}$\,Gauss, this RRAT has timing properties similar
to many canonical radio pulsars (McLaughlin et al. 2006, 2009). The
source position has been refined through radio timing analysis of the
bursts: RA 13:17:46.26, Dec -57:59:30.32 (J2000) with a 1$\sigma$
error of 0\farcs57 (McLaughlin et al. 2009). X-ray observations resulted in
a luminosity upper limit on its X-ray counterpart of
L$_X<7.5\times10^{32}$\ergs\, (assuming a distance of 3.2\,kpc; Rea \&
McLaughlin 2008).

We observed this object using three infrared bands (J, H, K$_{\rm s}$). No infrared counterpart has been detected, with limiting magnitudes reported in Tab.\,1, and a 3$\sigma$ upper limit on the K$_{\rm s}$ luminosity of L$_{\rm NIR}<1.3\times10^{30}$\ergs\,. We show in Fig.\,\ref{map1317} the K$_{\rm s}$ finding chart around \rrc 's  radio timing position.
 

\begin{figure*}
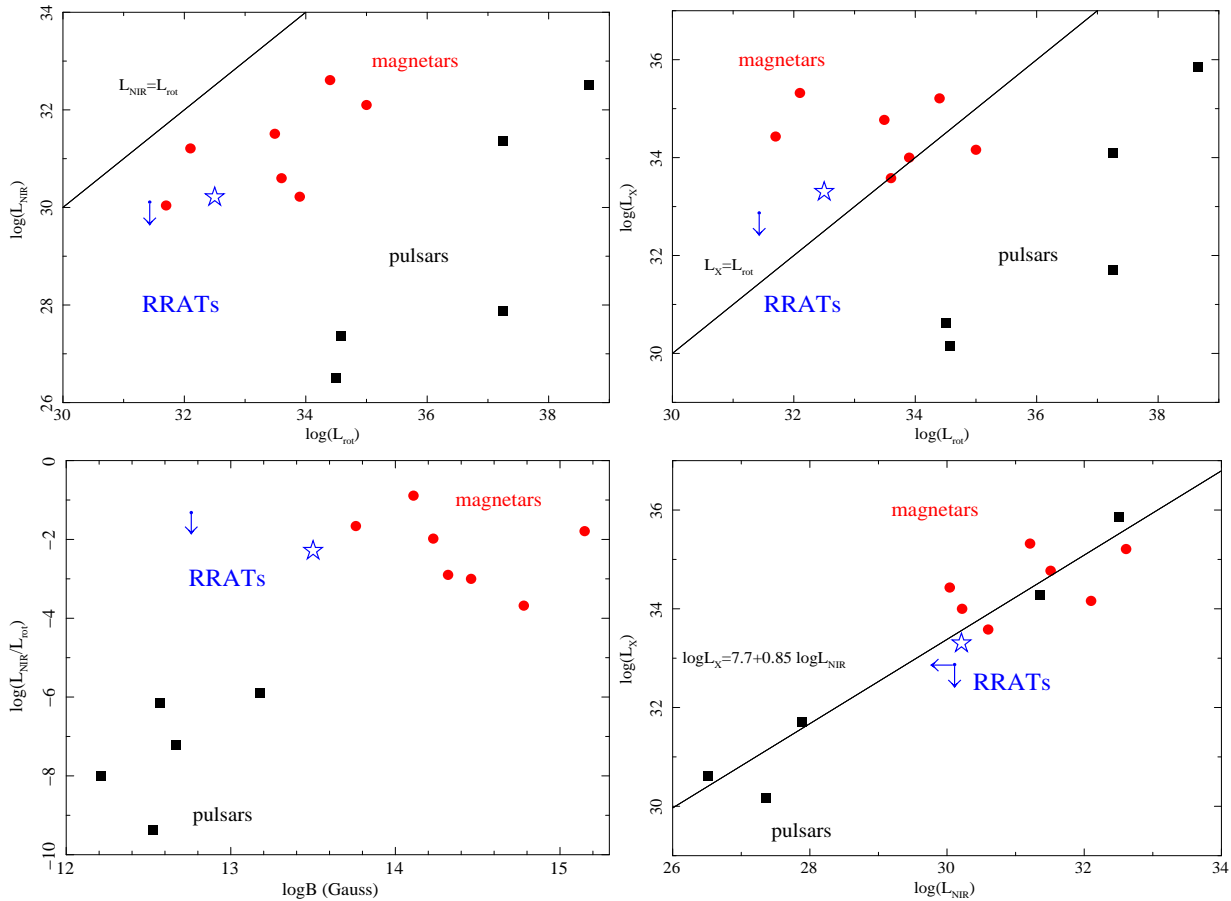

\vbox{
\hbox{
\psfig{figure=Lir_vs_Edot_last.ps,width=8cm,height=6cm,angle=270}
\psfig{figure=Lx_vs_Edot_last.ps,width=8cm,height=5.8cm,angle=270}}
\hbox{
\psfig{figure=LirEdot_vs_B_last.ps,width=8cm,height=6cm,angle=270}
\psfig{figure=Lx_vs_Lir_last.ps,width=8.3cm,height=5.85cm,angle=270}}}
\caption{Squares, circles and the star indicate rotation-powered pulsars, magnetars, and \rra, respectively, while the arrow shows \rrc's upper limit (see also Tab.\,3). {\em Top panels}: infrared and X-ray luminosities as a function of the rotational luminosity (solid lines represent L$_{\rm NIR}$=L$_{\rm rot}$ and L$_{\rm X}$=L$_{\rm rot}$, respectively). {\em Bottom panels}:  on the left we report on the infrared and rotational luminosity ratio (L$_{\rm NIR}$/L$_{\rm rot}$) versus the pulsar magnetic field, while on the right-side we show the X-ray versus the infrared luminosities. The solid line here represents the fit to the data (see text for details). For all the objects  L$_{\rm NIR}$ and L$_{\rm X}$ have been computed in the K$_{\rm s}$ and 1-10\,keV bands, respectively. Errors on the source distances are not taken into account in the luminosity uncertainties.}
\label{plots}
\end{figure*}


\section{Discussion}
\label{discussion}

Optical counterparts have been identified so far only for ten
rotational powered radio pulsars, and only five of them were observed in the infrared band (see Mignani et al. 2009 for a recent
review). The radio pulsars' infrared/optical emission depends mainly on
the pulsar age: young objects show pure magnetospheric emission,
middle-aged ones appear to have composite spectra with an additional
thermal component most probably arising from the cooling neutron star
surface, while for older objects there is more confusion, but
on average there is evidence for a dominant magnetospheric component
(Mignani et al.  2001; Zharikov et al. 2004). \\ 
\indent 
The situation is more favorable in the case of XDINS. Four XDINSs have
secure optical identifications (V$\sim 25$; van~Kerkwijk \& Kaplan 2007), with a thermal optical emission which
exceeds by a factor of $\sim 10$ the extrapolation of their
blackbody-like soft X-ray spectrum.  This suggests that the thermal
component responsible for the optical emission arises from a cooler
and larger area, with respect to their thermal X-ray emission. None of them have been detected
in the infrared band yet (Lo Curto et al. 2007). \\ 
\indent 
The neutron star infrared zoo, has been joined some years ago by the
magnetars. In particular, observations from large telescopes such as i.e. ESO, Gemini, CFHT, and Keck, led to the discovery of faint (K$_{\rm s}\sim$20) and, in some cases
variable, infrared counterparts for many magnetars (see
Tab.\,3 for a complete list, and also Israel et al. 2004; Testa et al. 2008; Mereghetti 2008). It is not yet clear whether infrared emission in
magnetars arise from reprocessed X-ray emission via a fossil disk
around the neutron star (Chatterjee et al. 2000; Perna et al. 2000),
or of magnetospheric origin (Beloborodov \& Thompson 2007), but it is
certainly true that with respect to normal radio pulsars, their infrared emission is much more efficient (see also below and
Fig.\,\ref{plots}). Other crucial peculiarities of magnetars' infrared
emission are their strange infrared colors, usually very red with
respect to their field stars (see e.g. Israel et al. 2004).


\begin{table*}
\centering{
\begin{tabular}{lccccccr}
\hline
\hline
Source & d(kpc) &  log$(L_{\rm rot})$ & log(L$_{\rm NIR}$) & log(L$_{\rm X}$)  & log(B) & Age $\tau$ (kyr) & Refs. \tabularnewline
\hline  
Crab & 1.7 & 38.66   &   Ê32.51 & 35.85 & 12.57 &  1.24 & 1,2,3 \tabularnewline
PSR\,1509--58 & 5.2 & 37.25  &   31.36 Ê & 34.28 & 13.18 &  154&  4,5,6 \tabularnewline
Vela & 0.3 & 37.25  &  27.88  & 31.70 & 12.53 & 11.3 & 7,8,9,10 \tabularnewline
PSR\,0656$+$14  & 0.3 & 34.58  &  27.36  &  30.16 & 12.67 & 111 & 11,12,13,14 \tabularnewline
Geminga & 0.16 & 34.50  &  26.51  & 30.62 & 12.21 & 342 & 11,15,16  \tabularnewline
\hline
1E\,1547--5408$^{*}$ & 9.0 & 35.00 &  32.10 &  34.16 & 14.32 &  1.41 & 17,18,19 \tabularnewline
SGR\,1806--20$^{*}$  &  8.7 & 34.40 & 32.61 & 35.21 & 15.15 &  0.22 & 20,21,22 \tabularnewline
1E\,1048--5937$^{*}$  & 3.0 & 33.90 &  30.22 &  34.00 & 14.78 & 2.68 & 23,24,25 \tabularnewline
XTE\,1810--197$^{*}$  & 4.0 & 33.60 &  30.60 &  33.58 & 14.46 &  1.13 & 26,27 \tabularnewline
SGR\,0501$+$4516$^{*}$  & 5.0 & 33.49 & 31.51 & 34.77  & 14.23 & 1.32 & 28,29 \tabularnewline
4U\,0142$+$61$^{*}$  & 5.0 & 32.10 &  31.21 &  35.32 & 14.11 &  67.7 & 30,31 \tabularnewline
1E\,2259$+$586$^{*}$  & 3.0 & 31.70 &  30.04 &  34.43 & 13.76  &  228 & 32,33,34,35\tabularnewline
\hline
RRAT\,J1819--1458 & 3.6 & 32.50 &  30.21 ($<$30.18) & 33.30  & 13.50 & 120 & this work, 36, 37\tabularnewline
RRAT\,J1317--5759 & 3.2 & 31.43 &  $<$30.11 & $<$32.87  & 12.76 & 3330 & this work, 36, 38 \tabularnewline

\hline
\hline
\end{tabular}
\label{tabplots}
\caption{Characteristics of pulsars and magnetars detected in the K$_{\rm s}$ near-infrared band, compared to the candidate infrared counterpart to \rra\, and \rrc. Luminosities are in units of \ergs. The latters are calculated in the K$_{\rm s}$-band and in the 1-10\,keV energy range for  L$_{\rm NIR}$ and L$_{\rm X}$, respectively.  The magnetic field values (B) are reported in Gauss. $^{*}$ Variable in the X-ray and NIR bands, not necessarily in a correlated fashion.  (1) Sollerman  (2003); 
(2) Sollerman et al (2000);
(3) Weisskopf et al. (2000);
(4) Kaplan \& Moon (2006);
(5) Gaensler et al. (1999); 
(6) DeLaney et al. (2006);
(7) Shibanov   et al. (2003); 
(8) Dodson et al. (2003);
(9) Mignani et al. (2003); 
(10) Pavlov et al. (2001);
(11) Koptsevich et al. (2001); 
(12) Brisken et al. (2003); 
(13) Pavlov  et al. (1997); 
(14) Shearer et al. (1997)
(15) Caraveo et al. (1996); 
(16) Kargaltsev et al. (2005); 
(17) Israel et al. (2009);
(18) Camilo et al. (2007);
(19) Bernardini et al. (2010 in prep); 
(20) Israel et al. (2005);
(21) Bibby et al. (2008)
(22) Eikenberry et al. (2004);
(23) Wang \& Chakrabarty (2002);
(24) Mereghetti et al. (2004);
(25) Gaensler et al. (2005);
(26) Israel et al. (2003);
(27) Rea et al. (2004);
(28) Levan et al. (2010 in prep);
(29) Rea et al. (2009b);
(30) Hulleman et al. (2004); 
(31) Rea et al. (2007); 
(32) Hulleman et al. (2001); 
(33) Kothes et al. (2002);
(34) Woods et al. (2004);
(35) Tian, Leahy \& Li 2010;
(36) McLaughlin et al. (2006);
(37) McLaughlin et al. (2007);
(38) Rea \& McLaughlin (2008).}}
\end{table*} 



We report in this paper on the first infrared observations of RRATs, performed using the NACO camera on the ESO-VLT (see
\S\ref{data}).  We found that among the two studied RRATs, \rra\, and
\rrc, only the former has a candidate infrared counterpart (see
Fig.\,\ref{map1819}) within our limiting magnitudes (see Tab.\,\ref{tabobs}).  Unfortunately,
we could detect the possible counterpart only in one K$_{\rm s}$ pointing
(the deepest we had), with a magnitude of K$_{\rm s}=20.96\pm0.10$,
hence we have no handle on the colors of this source because our J and
H observations were not deep enough (see Fig.\,3 for the limits on its
color). On the other hand we have also no handle on the possible
variability of this infrared counterpart (a behavior typical of
magnetar infrared emission; Israel et al. 2002; Rea et al. 2004; Israel et al. 2009), this is why we consider it as a candidate
counterpart which needs to be confirmed with further data.\\
\indent
With a magnitude of $K_{\rm s}=20.4$ (after correcting for
the extinction in the $K_{\rm s}$-band derived from the X-ray
absorption value N$_{\rm H}\sim9\times10^{21}$\cm2; Predehl \& Smith 1995; Cardelli, Clayton \& Mathis 1989), our candidate source A has a K$_{\rm s}$ flux density of
$F_{K_{\rm s}}\sim4.6\times10^{-29}$
erg\,s$^{-1}$\,cm$^{-2}$\,Hz$^{-1}$. If confirmed, this relatively
bright infrared emission strengthens the connection of \rra\, with the
magnetar class. In particular, the intensity of its candidate
counterpart would place it well among the magnetar luminosities
compared with those of normal pulsars.  As for the magnetars, the luminosity of this possible counterpart, lies above the extrapolation of the blackbody modelling its X-ray emission (McLaughlin et al. 2007). In Tab.\,3 and
Fig.\,\ref{plots} we computed the near-infrared luminosities
(L$_{\rm NIR}$) in the K$_{\rm s}$ band of all isolated neutron stars detected at such wavelength (namely 5 radio pulsars and 7 magnetars). Furthermore, we calculated
their X-ray luminosities in the 1-10\,keV band\footnote{We used for
  all magnetars the fluxes resulting from a spectral fit with a
  resonant cyclotron scattering model or two blackbodies (Rea et al. 2008, 2009b), with the aim
  of not overestimating the N$_{\rm H}$ values, which would have
  spuriously produced overestimated X-ray luminosities.} taking care
of choosing X-ray observations as close as possible to the date of the
near-infrared ones, crucial for most of the magnetars which show
variable emission in both such energy bands. However, given the
non-exact simultaneity of the X-ray and near-infrared observations
this luminosity comparison should be taken with the due
uncertainty. \\ 
\indent 
In Fig.\,\ref{plots} (top panels) we plot the infrared and X-ray luminosities of pulsars, magnetars and the two RRATs studied in this paper, as a function of their rotational energies (see also Tab.\,3).  It is clear that while  the rotational energy is not at all sufficient to power magnetars' and \rra\, X-ray emission, it might well suffice to be the reservoir for their infrared luminosity, as the radio pulsar case. However, even if powered by a similar mechanism, the efficiency of magnetars in converting rotational power in infrared luminosity, is much higher than normal radio pulsars, possibly due to their high magnetic fields (see also Mignani et al. 2007, 2009; Zane et al. 2008). This is particularly clear in Fig.\,\ref{plots} (bottom-left panel), where we see how this efficiency in much higher for highly magnetic sources (as \rra). \\
\indent
Pulsars detected in the near-infrared band are also
observed as X-ray emitters: this is not only due to the trivial
conclusion of being the closest, hence the most favorable for
detecting their faint multi-band emission, but it is also an intrinsic
connection. In particular, young objects as these particular pulsars,
present more powerful magnetospheric emission, responsible for both
the near-infrared and X-ray emission, which of course combined with
their close distances, make them visible in both bands. Comparing the near infrared and X-ray luminosity of all isolated neutron stars detected in both these bands, we found an intriguing correlation (see Fig.\,\ref{plots}  bottom-right panel):
$ log(L_{\rm X}) = 7.7\pm2.8 + (0.85\pm0.09)\times log(L_{\rm NIR}) .$ This empirical fit resulted in a $\chi^2=1.3$, with the magnetars having the largest spread over the model. This relation might be use to have a very rough estimate of the infrared luminosity of a known or new isolated neutron star, however, given the large errors in the distances (which have not been taken into account in the fit), this relation should be used with the due uncertainty.
\\ 
\indent 
In this picture, the K$_{\rm s}$ magnitude of our proposed candidate
counterpart to \rra\, (source A) is too bright to be the counterpart of a
normal radio pulsar, given its low rotational power, while it is
well in line with the values observed for magnetars (see Fig.\,\ref{plots}  top and bottom-left panels). This
would represent only one among a series of other properties which
connects this RRAT to the magnetar class, as its inferred dipolar B
field ($5\times10^{13}$\,Gauss), its bright X-ray emission with
respect to other RRATs (McLaughlin et al. 2007; Rea \& McLaughlin
2008; Kaplan et al. 2009), as well as the presence of a diffuse X-ray
nebula proposed to be powered by the magnetic energy (Rea et
al. 2009a). \\ 
\indent

\section{Conclusion}
\label{conclusion}

Near-infrared observations of \rra\, and \rrc\, resulted in a possible candidate counterpart for the former (K$_{\rm s}=20.96\pm0.10$), while no infrared counterpart was detected for the latter (within the limiting magnitudes reported in Tab.\,1). However, the proposed counterpart to \rra\, needs further confirmation due to the relatively high chance coincidence probability. Studying the near-infrared and X-ray luminosities  of all isolated neutron stars detected in the K$_{\rm s}$-band, with respect to other characteristics such as the rotational luminosity and the magnetic field, we show that the rotational energy may well be the resevoir of the magnetars and \rra\, infrared emission. However, in this picture the efficiency of converting rotational energy in near-infrared luminosity should be necessarily  higher for highly magnetic sources.

\vspace{0.5cm}

Based on observations collected at the European Southern Observatory,
Paranal, Chile under programs ID: 077.D-0395(A), 077.D-0395(B),
077.D-0395(C), and 077.D-0395(D). We wish to thank the ESO staff for
the support during these observations, M. Gilllon who made available the deconvolution code DECPHOT, and the anonymous referee for his/her suggestions which improved this paper. NR is supported by a Ramon y Cajal Research Fellowship, and thanks R. Mignani for useful
discussion, and E. Keane for his comments on the manuscript. MAM is supported by a WV EPSCoR grant, and a SAO guest
investigator grant, and B.M.G. acknowledges the support of the Australian Research
Council through grant FF0561298.

\label{lastpage}

\end{document}